\documentclass[11pt]{PoS}
\pdfoutput=1
\usepackage[T1]{fontenc}
\usepackage[utf8]{inputenc}
\usepackage{amsmath,amssymb,amsfonts}
\usepackage{multicol}
\usepackage{dcolumn,booktabs,colortbl}
\usepackage[leftcaption]{sidecap}
\usepackage{wasysym}
\usepackage{defs}
\title{Prospects and status of quark mass renormalization in three-flavour QCD}
\ShortTitle{Prospects and status of quark mass renormalization in $\nf=3$ QCD}
\author{\hfill\parbox{4cm}{\small\it%
CERN-PH-TH-2015-203\\
FTUAM-15-26\\
IFT-UAM/CSIC-15-089\\
}}
\author{%
I.~Campos,$^{a}$ %
\speaker{P.~Fritzsch},$^{b}$ %
C.~Pena,$^{b,c}$ %
D.~Preti,$^{b}$%
A.~Ramos,$^{d}$ and %
A.~Vladikas$^{e}$%
\vskip0.25em\\
\llap{$^a$} Instituto de Física de Cantabria - IFCA-CSIC, \\
Avda. de Los Castros s/n, 39005 Santander, Spain \\
\llap{$^b$} Instituto de Física Te\'orica UAM/CSIC,
Universidad Autónoma de Madrid, \\
C/ Nicolás Cabrera 13-15,
Cantoblanco, Madrid 28049  \\
\llap{$^c$} 
Departamento de Física Teórica, Universidad Autónoma de Madrid,\\
Cantoblanco, Madrid 28049   \\
\llap{$^d$}
PH-TH, CERN, \\
CH-1211 Geneva 23, Switzerland\\%
\llap{$^e$}
INFN, Sezione di Tor Vergata, c/o Dipartimento di Fisica, Università di Roma Tor Vergata, \\
Via della Ricerca Scientifica 1, I-00133 Rome, Italy
\vskip0.25em\\
E-mail:~\email{isabel.campos@csic.es}, \email{p.fritzsch@csic.es}, \email{carlos.pena@uam.es}, 
\email{david.preti@csic.es}, \email{alberto.ramos@cern.ch}, \email{vladikas@roma2.infn.it} %
}
\abstract{%
We present the current status of a revised strategy to compute the running of
renormalized quark masses in QCD with three flavours of massless O(a) improved
Wilson quarks. The strategy employed uses the standard finite-size scaling
method in the Schr{\"o}dinger functional and accommodates for the non-perturbative
scheme-switch which becomes necessary at intermediate renormalized couplings as
discussed in \href{http://arxiv.org/abs/1411.7648}{[1411.7648]}.
}
\FullConference{%
The 33rd International Symposium on Lattice Field Theory\\
14 -18 July 2015\\
Kobe International Conference Center, Kobe, Japan}
\begin{document}

\begin{figure}
    \centering
    \includegraphics[width=0.9\textwidth]{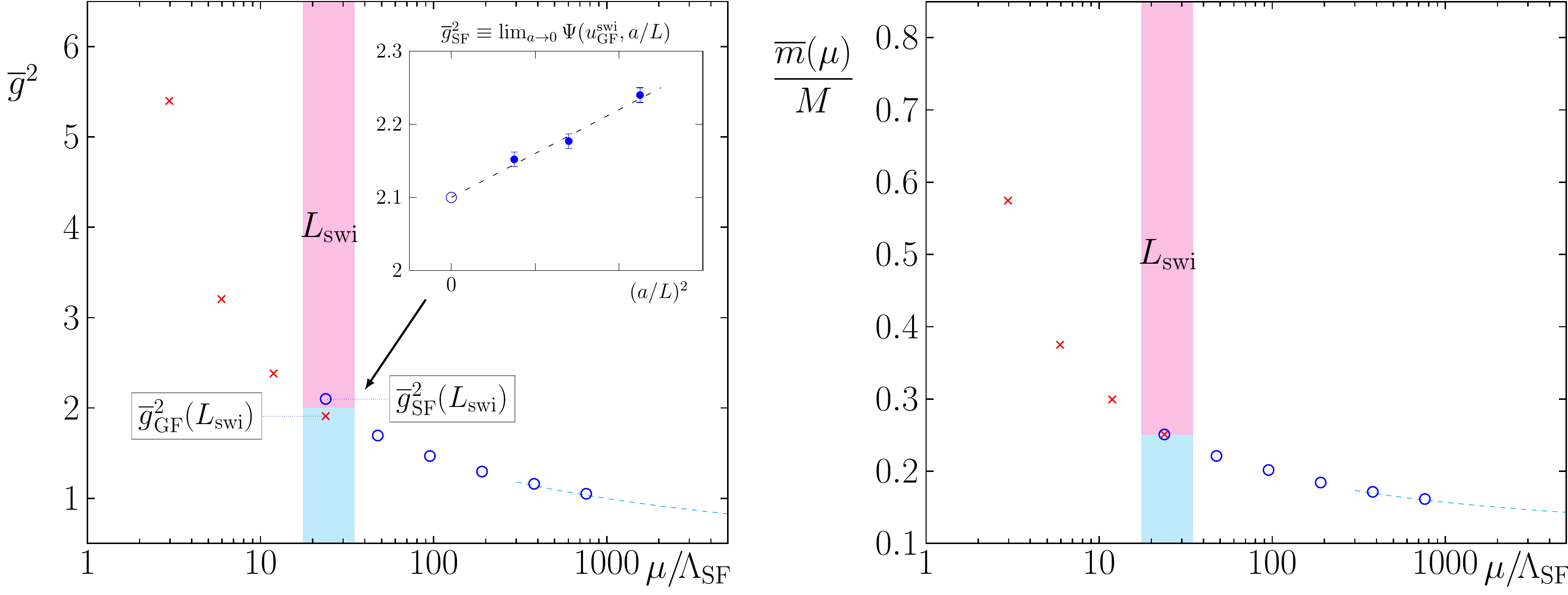}
    \caption{Sketch of the overall strategy to determine the $\Lambda$-parameter  
             in $\nf=3$ QCD (\textit{left}) as well as the renormalization group 
             running to invariant quark mass $M$ (\textit{right}).
            }
    \label{fig:sketch}
\end{figure}

\section{Introduction}

In a mass-independent renormalization scheme for QCD, the renormalization group
(RG) equations for the running coupling and quark mass read
\begin{align}
         \mu\dfrac{\partial}{\partial \mu}\gbar(\mu) =\beta(\gbar)  &\qquad \gsim \qquad -\gbar^3({ b_0}+{ b_1}\gbsq+{ b_2}\gbar^4+\ldots) \;, \label{eq:RGg2} \\[0.15em]
         \mu\dfrac{\partial}{\partial \mu}\mbar(\mu) =\tau (\gbar)  &\qquad \gsim \qquad -\gbar^2({ d_0}+{ d_1}\gbsq+\ldots)   \;,  \label{eq:RGm}
\end{align}
with universal coefficients $b_0, b_1, d_0$ and higher order ones, $b_{i>1}$
and $d_{j>0}$, which are scheme-dependent. In such a scheme it is natural to
first solve the RG equation for the coupling~\eqref{eq:RGg2} and then for the
mass~\eqref{eq:RGm}, as the latter depends parametrically on the coupling only.
Formally, the solutions can be exactly written down via renormalization group
invariants (RGI),
\begin{align}  \label{eq:Lambda}
   {\Lambda} &\equiv  
                         \mu\big[{b_0\gbsq(\mu)}\big]^{-{b_1}/({2b_0^2})}\,\ee^{-1/({2b_0\gbsq(\mu)})} 
                         \exp\bigg\{\!\!-\!\!\int_{0}^{\gbar^{\vphantom{A}}(\mu)}\!\dd{g}
                         \bigg[\dfrac{1}{\beta(g)}+\dfrac{1}{b_0g^3} -\dfrac{b_1}{b_0^2g}\bigg]\bigg\}  \;, \\[0.5em]
      { M_i} &\equiv  
                         \mbar_i(\mu)\big[{2b_0\gbsq(\mu)}\big]^{-{d_0}/({2b_0})}\,
                         \exp\bigg\{\!\!-\!\!\int_{0}^{\gbar^{\vphantom{A}}(\mu)}\!\dd{g}
                         \bigg[\dfrac{\tau(g)}{\beta(g)}-\dfrac{d_0}{b_0g}\bigg]\bigg\}  \;, \quad i\in\{\rm u,d,s,c,b,t\} \,, \label{eq:Mi}
\end{align}
which encode all information about the respective fundamental parameters of
QCD.  They are defined without relying on perturbation theory, and, while the
Lambda-parameter is trivially scheme-dependent, the RGI mass is independent
of the renormalization scheme.  Accordingly, the RGI parameters allow for easy
conversion (at high but finite $\mu$) between renormalized masses or couplings
in different (massless) schemes. The only remnant and non-trivial dependence is
on the number of quark flavours $\Nf$. The RGIs $\{\Lambda,M_i\}_{\Nf}$ have
been computed by the ALPHA collaboration in the past for $\Nf=0,2$ QCD, cf.
Refs.~\cite{Luscher:1993gh,DellaMorte:2004bc,Fritzsch:2012wq,Capitani:1998mq,DellaMorte:2005kg}.
In line with the current Coordinated Lattice Simulation effort
(CLS)~\cite{Bruno:2014jqa}, a new computation has been started to determine the
fundamental parameters  $\{\Lambda,M_i\}_{\Nf}$ for QCD with $\Nf=3$ dynamical
flavours. The standard techniques involved here have been refined over the years.
As intermediate mass-independent renormalization scheme the Schr\"odinger
functional (SF) is used with non-perturbatively improved Wilson fermions.
In the SF the renormalization scale is given by the inverse box size $\mu=1/L$.
Combined with a recursive finite-size scaling $L\to L/s$ in the continuum,
typically with $s=2$, one can reach a scale difference of $\Or(10^3)$ after
$N=10$ steps, and thus bridge the gap between typical large volume simulations
$(L\sim 6\,\fm)$ --- where the scale has been set --- and the regime where
perturbation theory can be safely applied.

\section{Revised strategy for $\Nf=3$}

In order to achieve a precision on $\alpha_{\rm s}(M_{\rm Z})$ that is better
than the current PGD estimate, the ALPHA collaboration has elaborated a new
strategy which minimizes statistical and systematic effects by combining the
standard SF coupling scheme with the new gradient flow (GF) sheme in the
Schr\"odinger functional~\cite{Fritzsch:2013je,Ramos:2015baa}. To this end one
has to match the two schemes non-perturbatively at some intermediate scale
$\mu_{\rm swi}=1/L_{\rm swi}$, as indicated in the sketch shown in the left
panel of Fig.~\ref{fig:sketch} and explained in more detail at last year's
Lattice conference~\cite{Brida:2014joa}. A status report has been given this
year~\cite{Sint:Lat15}. In Fig.~\ref{fig:sketch} the red crosses indicate
values for the gradient flow coupling $\ugf=\gbgf^2(L)$, $L\ge\lswi$, which is
used to make contact between a hadronic scale in large volumes, made available
by CLS, and the switching scale $\lswi$. From there on the Schr\"odinger
functional coupling $\usf=\gbsf^2(L)$, $L\le \lswi$, is used (blue circles) to
make contact with perturbation theory (dashed curve) at very high energy in
order to extract the $\Lambda$-parameter in physical units according to
eq.~\eqref{eq:RGg2}.

\begin{table}[t]
   \small\centering
   \begin{tabular}{LLLLLLLLLLL}\toprule
   \usf   & L/a  &  \beta       &  \kapcsf         &   \usf_\text{fit} \\\midrule
   1.1100 &  6   &  8.5403(55)  &  0.1323361(12)   &   1.1100(12)   \\
   1.1100 &  8   &  8.7325(72)  &  0.1321338(13)   &   1.1100(15)   \\
   1.1100 & 12   &  8.995(11)   &  0.1318617(10)   &   1.1100(24)   \\\midrule 
   1.4808 &  6   &  7.2618(28)  &  0.1339337(13)   &   1.4808(11)   \\
   1.4808 &  8   &  7.4424(38)  &  0.1336745(11)   &   1.4808(15)   \\
   1.4808 & 12   &  7.7299(89)  &  0.13326299(69)  &   1.4808(35)   \\\midrule 
   2.0120 &  6   &  6.2735(44)  &  0.1355713(17)   &   2.0120(32)   \\
   2.0120 &  8   &  6.4680(51)  &  0.1352362(15)   &   2.0120(39)   \\
   2.0120 & 12   &  6.7299(68)  &  0.1347591(10)   &   2.0120(49)   \\\bottomrule 
   \end{tabular}
   \caption{Simulation parameters and their known accuracy.}
   \label{tab:parms}
\end{table}

As aforementioned, solving the RG equation for the strong coupling is a
prerequesite for solving other RG equations such as the one for the mass which
we want to consider now. The general strategy for computing RGI quark masses
follows the decomposition~\cite{Capitani:1998mq,DellaMorte:2005kg}
\begin{align}\label{eq:RGI-def1}
    M_{i} &=  \frac{M}{\mbar(\muhad)} \times \mbar_i(\muhad) \;, & 
   \muhad &= 1/\lhad \;,
\end{align}
where $\lhad$ is some hadronic scale of $\Or(1\,\fm)$ and the total RG running
factor for the mass in the continuum, ${M}/{\mbar(\muhad)}$, connects
the renormalized current quark mass $\mbar_i(\muhad)$ to its RGI value.
This factor does not depend on the individual quark flavour $i\in\{{\rm
u/d},{\rm s},{\rm c},{\rm b}\}$, and in the $\Nf=2$ case, with
$\Delta[M/\mbar]=1.1\%$, it has been one of the dominant sources of errors in
the determination of $M_{\rm s}$~\cite{Fritzsch:2012wq} and $M_{\rm
b}$~\cite{Bernardoni:2013xba}. Accordingly, it is advisable to reduce this
error systematically as much as possible in the forthcoming determination. 
Due to the recursive step-scaling procedure the total RG factor decomposes
into
\begin{align}
        \dfrac{M}{\mbar(\muhad)} &= \dfrac{M}{\mbar(\mupt)}  \times \prod^{N}_{i=1} {\sigP(u_i)}   \;,
        &   u_{i} &= \gbsq(\lhad/2^i) \,,
        &   \mupt &= 2^{N} \muhad \,,
\end{align}
i.e., a product of step-scaling functions (SSF), $\sigP(u_i)$, and a part that can
be safely evaluated in perturbation theory using eq.~\eqref{eq:Mi} after 
a sufficiently high scale $\mupt$ is reached. The SSF reads 
\begin{align}
        {\sigP(u_i)} &= \exp\!\bigg[ -\!\int_{\gbar(\mu_i)}^{\gbar(\mu_i/2)} \!\!\!\!\dd{g}\, \dfrac{\tau(g)}{\beta(g)} \bigg]_{\gbsq(\mu_i)=u_i}^{m_{q}=0}
        \hspace*{-1em}= \lim_{a\to 0} \SigP(u_i,a/L) \;, &
  {\SigP(u_i,a/L)} &= \dfrac{\ZP(g_0,2L/a)}{\ZP(g_0,L/a)}   \;,
\end{align}
i.e., it is obtained from its lattice SSF $\SigP$ along a line of constant
physics (LCP) when $a\to 0$. This is achieved by implicitly prescribing a value
$u_i$ to the renormalized coupling $\gbsq$ in a given scheme at vanishing quark
mass $m_q$. The relevant scale-dependent renormalization constant $\ZP$ is
obtained from standard correlation functions ($\fP,\fone$) in the SF
by the \emph{renormalization condition}
\begin{align}
    \left[ {\ZP(g_0,L/a)}\frac{\fP(L/2)}{\sqrt{3\fone}} \right]_{m_q=0}^{\theta} &= c_3(a/L) \;,
\end{align}
employing the known tree-level normalization $c_3(a/L)$ with vanishing boundary gauge
fields and $(T/L,\theta,m_q)=(1,0.5,0)$,
cf.~\cite{Capitani:1998mq,DellaMorte:2005kg}. For the three-flavour computation
we are following as closely as possible the strategy for the running coupling,
meaning that we also switch the scheme for setting up the LCP above/below
$\lswi$.
The computation of the total RG factor is easily extended to this
two-scheme case (GF/SF coupling) and reads
\begin{subequations}\label{eq:2-scheme-run}
\begin{align}
    \dfrac{M}{\mbar(\muhad)} &= \dfrac{M}{\mbar(\mupt)}  \times \dfrac{\mbar(\mupt)}{\mbar(\muswi)} \times \dfrac{\mbar(\muswi)}{\mbar(\muhad)}   \;,  
    &   \mupt &= 2^{N_{\rm SF}} \muswi \,, &   \muswi &= 2^{N_{\rm GF}} \muhad \,,  
\end{align}
with
\begin{align}
    \dfrac{\mbar(\mupt)}{\mbar(\muswi)} &= \prod^{N_{\rm SF}}_{i=1}  {\sigP(\usf_i)}   \;, 
    &   \usf_{i} &= \gbsf^2(\lswi/2^i) \,,  \\
    \dfrac{\mbar(\muswi)}{\mbar(\muhad)} &= \prod^{N_{\rm GF}}_{i=1} {\sigP(\ugf_i)}   \;, 
    &   \ugf_{i} &= \gbgf^2(\lhad/2^i) \,.
\end{align}
\end{subequations}
By doing so we directly profit from the running coupling
project~\cite{Brida:2014joa,Sint:Lat15} as the tuning to vanishing quark mass
and fixed coupling has already been done very accurately in terms of the
bare parameters $(\beta,\kapc)$ of the theory. An example for the parameters in
use is given for the SF coupling part in Table~\ref{tab:parms}. Fixing the SF
coupling to the values in the first column allows for individual $L/a$ to
use the known parametrization $\usf_{\rm fit}(\beta)$ to determine the relevant
value of $\beta=6/g_0^2$, as given with error in column 3. Our knowledge of the
critical line $\kapcsf(\beta)$, where $m_q=0$, subsequently gives the critical
hopping parameters as listed in column 4. Column 5 finally quotes the value of 
$\usf_{\rm fit}(\beta)$ at the previously determined value of $\beta$ and thus
reflects our knowledge of the SF coupling.

\section{Preliminary results}

\begin{SCfigure}
  \small\centering
  \includegraphics[height=8cm]{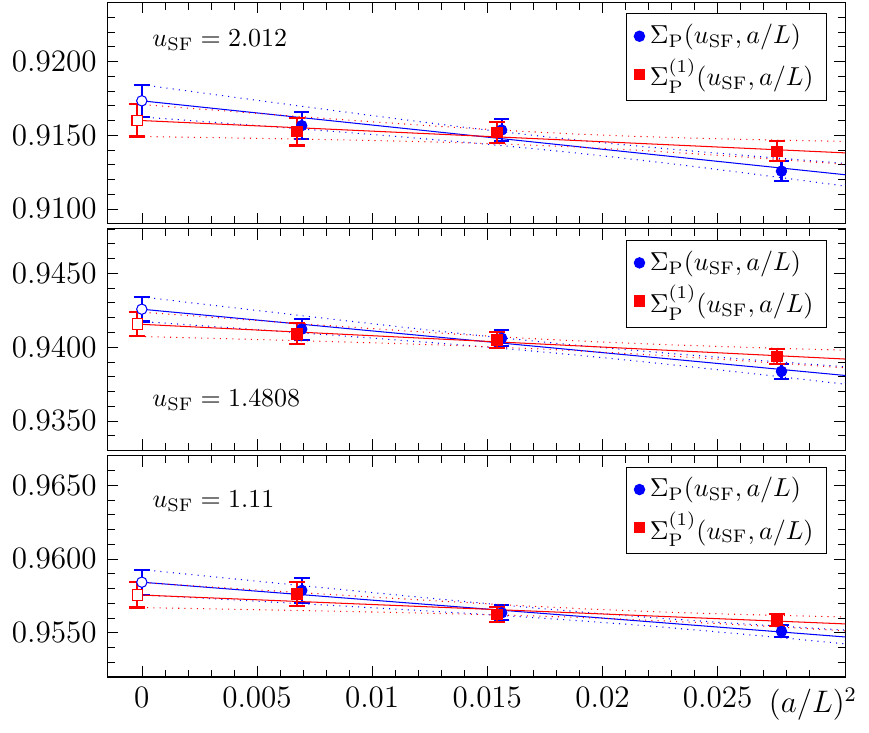}
  \caption{Individual (preliminary) continuum extrapolations of
           $\SigP(\usf,a/L)$ and its one-loop improved results
           $\SigP^{(1)}(\usf,a/L)$ using all available lattices
           at fixed SF coupling $\usf\in\{2.012,1.4808,1.110\}$.
          }
  \label{fig:CL}
\end{SCfigure}

\begin{figure}[t]
  \small\centering
  \includegraphics[width=0.49\textwidth]{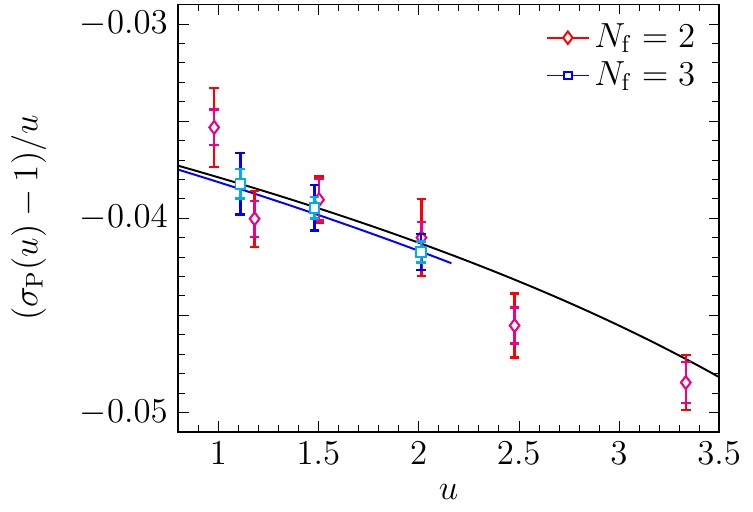}
  \includegraphics[width=0.49\textwidth]{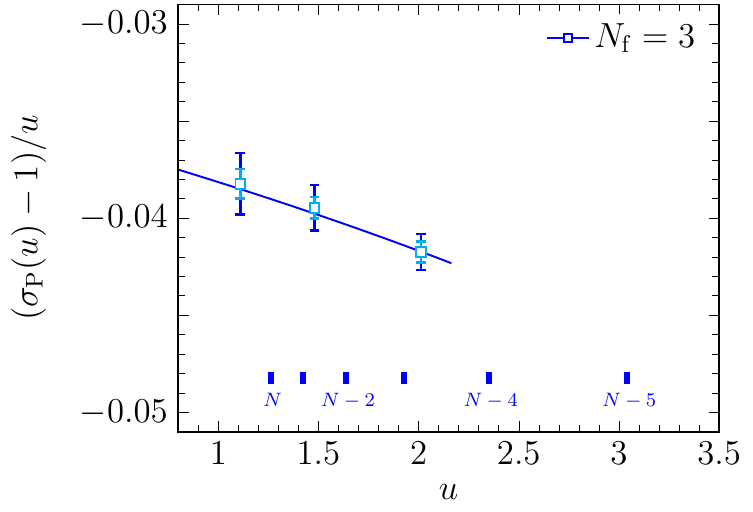}
  \vskip-1em
  \caption{Comparison of step-scaling functions $\sigP(\usf)$ for $\Nf=2$ (red
           diamonds) with our preliminary results in the three-flavour theory (blue 
           squares). The solid lines represent the perturbative running
           using the 2-loop $\tau$-function and 3-loop $\beta$-function
           in the 2- and 3-flavour theory, respectively.
          }
  \label{fig:mass-SSF}
\end{figure} 

In the $\Nf=3$ running coupling project~\cite{Brida:2014joa,Sint:Lat15}, the
step-scaling function (SSF) for the RG running of the SF coupling in the
high-energy regime, $\sigsf(\usf)$ $(2.012\ge \usf\ge 1.110)$, has been
finished.  Its determination requires field configurations with non-vanishing
boundary conditions on the gauge field~\cite{Luscher:1992an}, while determining
renormalization factors, such as $\ZP$ and thus $\SigP$, usually proceeds with
vanishing boundary gauge fields. To this end we have started production of
$\Nf=3$ field ensembles for $L/a\in\{6,8,12\}$ and their $2L/a$ counterparts.
As for the step-scaling function $\sigsf$ this is done at 8 values of $\usf$
covering the aforementioned range. The bare parameters for these runs are
summarized in Table~\ref{tab:parms} for the three values
$\usf\in\{2.012,1.4808,1.110\}$ at which we already have results on the
computationally more expensive $2L/a=24$ lattices.
We present preliminary results for the lattice step-scaling $\SigP(u,a/L)$ in 
Figure~\ref{fig:CL} (blue filled circles), together with its 1-loop 
improved version (red filled squares), defined by
\begin{align}\label{eq:SigP-1lp-impr}
       \SigP^{(1)}(u,a/L) &= \dfrac{\SigP(u,a/L)}{1+\delta_{\rm P}(a/L) \, u}  \;.
\end{align}
This effectively reduces the leading cutoff effects in the lattice step-scaling
function $\SigP(u,a/L)$. The coefficients $\delta_{\rm P}(a/L)$ can be inferred
from~\cite{Sint:1998iq}. Although, it is not clear whether the rightmost points
for $\SigP(u,1/6)$ are still in the region where $(a/L)^{n>2}$ cutoff effects
are negligible, cf.~\cite{DellaMorte:2005kg}, for the time being we use all 
three data points and perform the continuum extrapolation using the ansatz 
$\SigP(u,a/L) =  A + B \, (a/L)^2$ to determine $A \equiv \sigP(u)$. We obtain
\begin{center}
  \begin{tabular}{llll}\toprule
          $u                                 $ &  $ 2.012         $ & $ 1.4808       $  & $ 1.110         $ \\\midrule
          $\sigP(u) ~\text{from}~ \SigP      $ &  $ 0.9173(11)(19)$ & $ 0.9426(8)(17)$  & $ 0.9584(9)(17) $ \\
          $\sigP(u) ~\text{from}~ \SigP^{(1)}$ &  $ 0.9160(11)(8) $ & $ 0.9416(8)(9) $  & $ 0.9576(9)(9)  $ \\\bottomrule
  \end{tabular}
\end{center}
where the first error is statistical and the $2^{\text{nd}}$ is the difference
to the result of a 2-pt weighted average which neglects the $L/a=6$ data. Our
results at fixed $u$ agree well at the one-sigma level. As seen in
Figure~\ref{fig:CL}, cutoff effects are reduced in the continuum extrapolation
for $\SigP^{(1)}$, as to be expected. Accordingly, we can take the values in the 
third row as our preliminary results for $\sigP(\usf)$ and note that at present
the statistical, $\delta_{\rm stat}$, and systematic error, $\delta_{\rm sys}$,
are $\lesssim 1\permil$.
To get an impression about the quality of these results, we compare them to
results obtained for the mass SSF in the two-flavour case. To
this end we reproduce Fig.3 of~\cite{DellaMorte:2005kg} in the left panel of
Figure~\ref{fig:mass-SSF} above. We add to the $\Nf=2$ data (red diamonds) our 
present results (blue squares) and note that the inner error bar is the purely
statistical error, and the outer error bars include the linearly added systematic 
errors. The latter have been determined as discussed earlier. The two solid
lines represent the perturbatively known SSF $\sigP^{\rm PT}(u)$ to the highest
available order in PT, cf.~\cite{DellaMorte:2005kg}, for the $\Nf=2, 3$ flavour
cases respectively. The $\Nf=3$ data falls in line with the PT SSF at small
couplings in the SF scheme, best seen in the right panel of the same figure.
We furthermore indicate the distribution of the couplings
$\usf_{i}=\gbsf^{2}(\lmax/2^i)$ as they have been in the $\Nf=2$ 
case~\cite{Fritzsch:2012wq}.

The fact that the new data scatters less about the PT behaviour at small
couplings compared to $\Nf=2$, is most likely a result of the intensified and
more systematic effort to set up the LCP~\cite{Brida:2014joa} which has become
possible due to algorithmic and computational advancements during the last 10
years. Especially at the switching scale $(\usf=2.012)$, also the total error
could be reduced significantly. With 5 additional, uncorrelated continuum data
points to be added to this picture soon, $\usf_{i}\in (1.110,2.012)$, we will
be in the comfortable position to fit the non-perturbative SSF to much higher
accuracy than in the past.
However, to give an upper bound on the estimate of the total error of the full
$\Nf=3$ RG running mass factor right now, we can naively assume, using
$\Delta[\sigP] = \delta_{\rm stat} + \delta_{\rm sys} \le 2\permil$ as
discussed above, that
\begin{align}\label{eq:bound}
        \Delta\left[\frac{M}{\mbar(\muhad)}\right] &\simeq \prod_{i}^{N}\Delta[\sigP(u_{i})] \simeq \sqrt{N} \cdot 2\permil  \simeq
        \begin{cases}
                0.49 \%  & \text{for } N=6  \quad \text{(as for $\Nf=2$)} \\
                0.63 \%  & \text{for } N=10   \\
        \end{cases} \;.
\end{align}
This means that the total error compared to $\Nf=2$ will be reduced by at least
a factor of 2. Note that we have $N=N_{\rm SF}+N_{\rm GF}$ according
to~\eqref{eq:2-scheme-run}. The main assumption entering this upper bound is
that we control the error on $\sigP(u_{i})$ as good in the GF scheme running
part at lower energies as we have reported here for the SF scheme running mass
part.
Needless to say that the naive bound in~\eqref{eq:bound} can be further reduced
by increasing the statistics on the lattice SSF to reduce the statictical error
even further. But as the systematic error is of the same size as the
statistical, it would be better to add another lattice spacing such as $L/a=10$ or
$16$ in order to check and further reduce the systematic error in the
continuum extrapolation.

\section{Summary}

We have presented first results towards a computation of the full RG factor for
the running mass in three-flavour QCD by the ALPHA collaboration. This is a
necessary ingredient to determine renormalization group invariant quark masses
from $\Nf=3$ large-volume ensembles provided by CLS. Systematic errors due to
large scale differences are controlled by following a traditional finite-size
scaling approach, using the Schr\"odinger functional as intermediate
finite-volume renormalization scheme to non-perturbatively connect low- and
high-energy regimes. 

A gain in the overall precision is achieved by (a) exploiting very precise
results from the ongoing $\Nf=3$ running coupling project, incorporating a
non-perturbative scheme switch at intermediate energies, and (b) via
computational advances in algorithms and hardware over the last decade. Both
together will allow us to reach an unprecedented precision in the
non-perturbative step-scaling function of the mass, provided that equally good
results are obtained for the running at lower energies. However, to finally
quote an RGI quark mass in physical units, one needs to incorporate the
scale-setting procedure which introduces yet another uncertainty and
modifies~\eqref{eq:RGI-def1}~to 
\begin{align}\label{eq:RGI-def2}
        M_{i} &=  \frac{M}{\mbar(\muhad)} \times \lim_{a\to 0} \left[ \frac{a\mbar_i(\muhad)}{a\fhad} \right] \times \fhad \;, & 
   \muhad &= 1/\lhad \;.
\end{align}
Here $\fhad$ is any physical quantity that has been used to set the scale
on the CLS ensembles.

Together with improvements in the scale-setting procedure w.r.t. the
two-flavour case, we can remain optimistic to reach for instance an 
overall precision less than $1\%$ on $M_{\rm s}$ itself.

\section*{Acknowledgments}
\small
The simulations were performed on the Altamira HPC facility and the GALILEO
supercomputer at CINECA (INFN agreement). We thankfully acknowledge the
computer resources and technical support provided by the University of
Cantabria at IFCA and at CINECA.

\bibliography{mainbib}
\end{document}